\documentclass[aps,pra,showpacs,twocolumn,floatfix,superscriptaddress]{revtex4}
\usepackage{bm,dcolumn,graphicx,amsmath,amsfonts,amssymb}
\usepackage{epsf}
\begin{document}
\title{Accurate calculations of Sr properties for a high-accuracy optical clock}

\author{S. G. Porsev}
\email{sporsev@gmail.com}
\affiliation{JILA, National Institute of Standards and Technology
and University of Colorado, Department of Physics, University of
Colorado, Boulder, CO, 80309-0440, USA} \affiliation{Petersburg
Nuclear Physics Institute, Gatchina, Leningrad district, 188300,
Russia}

\author{Andrew D. Ludlow}
\altaffiliation{Present address:
NIST Time and Frequency Division, Boulder, Colorado, 80309, USA}

\author{Martin M. Boyd}

\author{Jun Ye}
\affiliation{JILA, National Institute of Standards and Technology
and University of Colorado, Department of Physics, University of
Colorado, Boulder, CO, 80309-0440, USA}

\date{\today}

\begin{abstract}
We have carried out calculations towards the goal of reducing the
inaccuracy of the Sr optical atomic clock to 1$\times$10$^{-17}$ and
below. We calculated a.c. polarizabilities of the $5s^2 \,\,
^1\!S_0$ and $5s5p \,\, ^3\!P_0^o$ clock states that are important
for reducing the uncertainty of blackbody radiation-induced
frequency shifts for the $^1\!S_0 - \,^3\!P_0^o$ clock transition.
We determined four low-lying even-parity states whose total
contribution to the static polarizability of the $^3\!P_0^o$ clock
state is at the level of 90\%. We show that if the contribution of
these states is experimentally known with 0.1\% accuracy, the same
accuracy can be achieved for the total polarizability of the
$^3\!P_0^o$ state. The corresponding uncertainty for the blackbody
shift at a fixed room temperature will be below 1$\times$10$^{-17}$.
The calculations are confirmed by a number of experimental
measurements on various Sr properties.
\end{abstract}

\maketitle
\section{Introduction}
The use of optical lattice-confined neutral atoms for the goal of
achieving a new level of time-keeping precision and accuracy has
become widely practiced (see, e.g.,~\cite{Katori03, ludlow06, Boyd2,
Campbell,letargat06,Baillard, takamoto06,LudZelCam08, NistYb2}). In
this scheme, ultracold atoms are confined in an optical lattice to
eliminate motion-related systematic frequency shifts. The lattice
laser wavelength is selected in such a way that the perturbation to
the clock frequency arising from a.c. Stark shifts induced by the
lattice laser for both clock states exactly cancel~\cite{Ye08}.

One indicator of clock performance is provided by the Allan
deviation characterizing the fractional frequency instability
$\sigma$.  For a signal-to-noise ratio given by the fundamental
quantum projection noise, the instability can be written as:
\begin{equation}
\sigma(\tau) \simeq  \frac{1}{Q} \frac{\sqrt{T}}{\sqrt{N \tau}}.
\label{sigma}
\end{equation}
Here $Q$ is the resonance quality factor defined as $Q= \nu/ \Delta
\nu$, where $\nu$ is the transition frequency and $\Delta \nu$ is
the linewidth. $N$ is the total number of particles measured in a
coherent time period of $T$, and $\tau$ is the total averaging time.
According to Eq.~(\ref{sigma}), atoms with the highest quality
factors $Q$ are preferred for a new generation of time and frequency
standards.

The highest $Q$'s are currently obtained for a narrow transition in
the optical domain~\cite{boyd06}. In particular, the highly
forbidden $^1\!S_0 \rightarrow \, ^3\!P_0^o$ transitions in divalent
atoms offer excellent possibilities for attaining a new level of
precision and accuracy for time-keeping. One of the most promising
candidates is $^{87}$Sr for which $\Delta \nu \approx$ 1.2
mHz~\cite{PorDer04,santra04, BoydPRA}, yielding a potential $Q >
10^{17}$. In a recent paper~\cite{LudZelCam08}, a systematic
uncertainty evaluation for a neutral Sr optical atomic standard was
reported at the 10$^{-16}$ fractional level, surpassing the best
evaluations of Cs fountain primary standards. The dominant
systematic frequency correction and uncertainty in that work arose
from the room temperature blackbody radiation (BBR). The fractional
frequency shift, $|\delta \nu_{BBR} /\nu_0|$, caused by the BBR is
proportional to the differential static polarizability of the two
clock states. For the $5s^2 \,\, ^1\!S_0 \rightarrow 5s5p \,\,
^3\!P_0^o$ transition in Sr, the BBR shift was calculated in
Ref.~\cite{PorDer06BBR} to be equal to 55.0(7)$\times 10^{-16}$. The
1\% uncertainty for the BBR shift originates mostly from
insufficient knowledge of the static polarizability of the $5s5p
\,\, ^3\!P_0^o$ state, with the most accurate calculation provided
in ~\cite{PorDer06BBR}. To further improve the Sr accuracy, it is
clear that better understanding of the Sr properties is needed to
give a more accurate determination of the BBR shift. The purpose of
this paper is to outline a clear path to achieve this goal.  To
improve the clock accuracy significantly, it is equally important
that a well-characterized homogeneous BBR environment surrounds the
Sr atoms in future experiments.

The improvement of the Sr clock accuracy requires a more accurate
determination of the differential static polarizability of the $5s^2
\,\, ^1\!S_0$ and $5s5p \,\, ^3\!P_0^o$ states. Note that the static
polarizability of the ground $5s^2 \,\, ^1\!S_0$ state is known at a
sufficiently low uncertainty $\sim$ 0.1\%~\cite{PorDer06BBR}. This
low uncertainty is made possible by a good knowledge of the matrix
element $|\langle 5s^2 \,\, ^1\!S_0 ||d|| 5s5p \,\, ^1\!P_1^o
\rangle|$~\cite{YasKisTak06}, where the intermediate state $5s5p
\,\, ^1\!P_1^o$ contributes to the polarizability of the $^1\!S_0$
state at the dominant level of 97\%. Consequently, the outstanding
challenge is to reduce the uncertainty of the polarizability of the
$^3\!P_0^o$ state to a similar level. Even sophisticated modern
relativistic methods of atomic calculations cannot provide such
accuracy. For this reason, a solution to this problem must combine
theoretical and experimental approaches. We show that four specific
intermediate states have a combined contribution to the total static
polarizability of the $5s5p \,\, ^3\!P_0^o$ state at the level of
90\%. When the contributions from these four states are determined
from experimental data at 0.1\% accuracy and the contributions of
all other discrete and continuum states are known at the level of
1-2\% from calculations, then the final 0.1\% uncertainty for the
polarizability of the $5s5p \,\, ^3\!P^o_0$ state will be achieved.
This strategy is the focus of this paper.

The paper is organized as follows. In Section II we briefly describe
the method of calculations. In particular, we discuss the
construction of basis sets and solving the multiparticle
Schr\"odinger equation. In Section III we discuss the blackbody
radiation effect and present the results of calculations for the
low-lying energy levels, a.c. polarizabilities, transition rates,
and oscillator strengths, and we analyze the results obtained.
Section IV contains concluding remarks. Atomic units ($\hbar = |e| =
m$ = 1) are used throughout the paper.

\section{Method of calculations}
The most complex problem in precise atomic calculations is
associated with the necessity to account for three types of electron
correlations, i.~e., valence-valence, core-valence, and core-core
correlations. The former are usually too strong to be treated
perturbatively, while the other two types of correlations cannot be
treated effectively with non-perturbative techniques, such as the
multi-configurational Hartree-Fock method~\cite{GraQui88} or the
configuration interaction (CI) method~\cite{KotTup87,JonFro94}.

Therefore, it is natural to combine the many-body perturbation
theory (MBPT) with one of the non-perturbative methods. In Ref.
\cite{DzuFlaKoz96b}, it was suggested to use MBPT in order to
construct an effective Hamiltonian for valence electrons. After
that, the multiparticle Schr\"odinger equation for valence electrons
is solved within the CI framework. Doing so allows us to find the
low-lying energy levels. Following the earlier works, we refer to
this approach as the CI+MBPT formalism.

In order to calculate other atomic observables, one has to construct
the corresponding effective operators for valence electrons
\cite{DzuKozPor98,PorRakKoz99P,PorRakKoz99J}. These operators
effectively account for the core-valence and core-core correlations.
In particular, to obtain an effective electric-dipole operator, we
solve random-phase approximation (RPA) equations, thus summing a
certain sequence of many-body diagrams to all orders of
MBPT~\cite{DzuKozPor98,KolJohSho82,JohKolHua83}. The RPA describes
shielding of an externally applied field by core electrons. Small
corrections due to, for instance, normalization and structural
radiation are omitted.

In the CI+MBPT approach, the energies and wave functions are
determined from the time-independent Schr\"odinger equation
$$H_{\rm eff}(E_n) \Phi_n = E_n \Phi_n,$$
where the effective Hamiltonian is defined as
$$H_{\rm eff}(E) = H_{\rm FC} + \Sigma(E).$$
Here $H_{\rm FC}$ is the Hamiltonian in the frozen core
approximation and $\Sigma$ is the energy-dependent correction, which
takes into account virtual core excitations. The operator $\Sigma$
completely accounts for the second-order perturbation theory over
residual Coulomb interaction. Determination of the second-order
corrections requires calculation of one- and two-electron diagrams.
The one-electron diagrams describe an attraction of a valence
electron by a (self)-induced core polarization. The two-electron
diagrams are specific for atoms with several valence electrons. The
number of the two-electron diagrams is very large and their
calculations are extremely time-consuming. In the higher orders the
calculation of two-electron diagrams becomes practically impossible.
Hence, it is more promising to account for the high-orders of the
MBPT indirectly. One of such methods was suggested in
Ref.~\cite{KozPor99O}, where it was shown that a proper choice of
the optimum initial approximation for the effective Hamiltonian can
substantially improve the agreement between calculated and
experimental spectra of many-electron atom.

We consider Sr as a two-electron atom with the core
[1$s$,...,4$p^6$]. The one-electron basis set for Sr includes
1$s$--14$s$, 2$p$--14$p$, 3$d$--13$d$, 4$f$--13$f$, and 5$g$--9$g$
orbitals, where the core- and 5$s$--7$s$, 5$p$--7$p$, and 4$d$--6$d$
orbitals are Dirac-Hartree-Fock (DHF) ones and all the rest are the
virtual orbitals. The orbitals 1$s$--5$s$ were constructed by
solving the DHF equations in the $V^N$ approximation, i.e. the core
and the 5$s$ orbitals were obtained from the DHF equations for a
neutral atom (we used the DHF computer code~\cite{BraDeiTup77}). The
6$s$--7$s$, 5$p$--7$p$, and 4$d$--6$d$ orbitals were obtained in the
$V^{N-1}$ approximation. That is, the 1$s$-�5$s$ orbitals
were``frozen'', one electron was transferred from the valence 5$s$
shell into one of the orbitals specified above, and the
corresponding one-electron wave function was found by solving the
HFD equations. We determined virtual orbitals using a recurrent
procedure similar to Ref.~\cite{BogZuk83} and described in detail
in~\cite{PorRakKoz99P,PorRakKoz99J}.

Configuration-interaction states were formed using
these one-particle basis sets. It is worth emphasizing that the
employed basis set was sufficiently large to obtain numerically
converged CI results. An extended basis set, used at the stage of
MBPT calculations, included 1$s$--21$s$, 2$p$--21$p$, 3$d$--20$d$,
4$f$--17$f$, and 5$g$--13$g$ orbitals.

\section{Results and discussion}

\subsection{Calculation of energies}
\begin{figure}[h]
\begin{center}
\includegraphics*[scale=0.35]{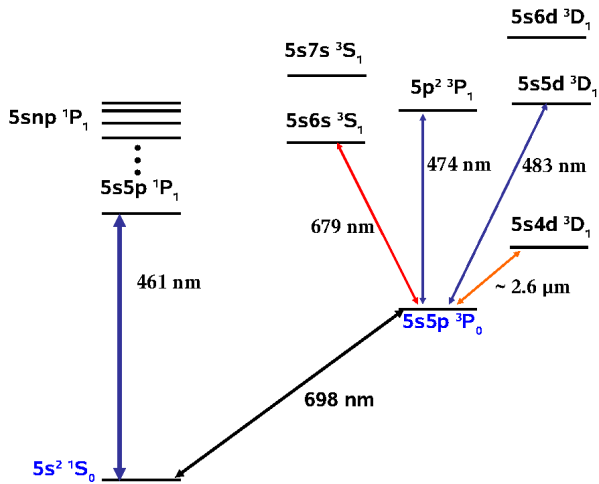}
\end{center}
\caption{(Color online) Low-lying energy levels and transition
wavelengths of atomic Sr, relevant for the determination of
polarizabilities of the $5s^2 \,\, ^1\!S_0$ and $5s5p \,\,
^3\!P_0^o$ clock states.} \label{Fig:levels}
\end{figure}

Solving the multiparticle Schr\"odinger equation we find low-lying
energy levels and their respective wave functions. In
Table~\ref{Energ} we present the calculated energies of the
low-lying states for Sr and compare them with experimental data. As
is seen from the table we focus mainly on the energy levels with
$J=1$. This is due to the fact that we are interested in calculation
of the electric dipole-dominated a.c. polarizabilities for the clock
states with total angular momentum $J=0$. Only intermediate states
with $J=1$ contribute to these polarizabilities. The energy level
diagram of these states is given in Fig.~\ref{Fig:levels}. The
energy levels were obtained in the framework of the conventional
configuration-interaction method as well as using the formalism of
CI combined with the many-body perturbation theory. Using the CI
method alone, the agreement of the calculated and experimental
energies is at the level of 5--10\%. The combination of CI and MBPT
improves the accuracy by approximately an order of magnitude.

\begin{table}
\caption{Energy differences (in cm$^{-1}$) with respect to the
ground $5s^2 \,\, ^1\!S_0$ state for the low-lying energy levels of
Sr.} \label{Energ}
\begin{ruledtabular}
\begin{tabular}{ccccc}
 Config. & Level
 & \multicolumn{1}{c}{\qquad CI}
 & \multicolumn{1}{c}{\qquad CI+MBPT}
 & \multicolumn{1}{c}{\qquad Experiment~\cite{Moo71}} \\
\hline
$5s4d$ & $^3D_1$   & 19571 & 18076 & 18159 \\
$5s4d$ & $^3D_2$   & 19587 & 18141 & 18219 \\
$5s4d$ & $^3D_3$   & 19617 & 18254 & 18319 \\
$5s4d$ & $^1D_2$   & 20166 & 19968 & 20150 \\
$5s6s$ & $^3S_1$   & 27488 & 29019 & 29039 \\
$5s5d$ & $^3D_1$   & 33358 & 35060 & 35007 \\
$5p^2$ & $^3P_1$   & 33511 & 35326 & 35194 \\
$5s7s$ & $^3S_1$   & 35695 & 37429 & 37425 \\
$5s6d$ & $^3D_1$   & 37985 & 39725 & 39686 \\
\hline
$5s5p$ & $^3P^o_0$ & 12490 & 14241 & 14317 \\
$5s5p$ & $^3P^o_1$ & 12663 & 14448 & 14504 \\
$5s5p$ & $^3P^o_2$ & 13022 & 14825 & 14899 \\
$5s5p$ & $^1P^o_1$ & 20832 & 21469 & 21698 \\
$5s6p$ & $^3P^o_1$ & 32110 & 33814 & 33868 \\
$5s6p$ & $^1P^o_1$ & 32487 & 34105 & 34098 \\
$4d5p$ & $^3D^o_1$ & 36699 & 36189 & 36264 \\
$4d5p$ & $^3P^o_1$ & 36944 & 37213 & 37303 \\
$5s7p$ & $^1P^o_1$ & 37275 & 38927 & 38907 \\
$5s7p$ & $^3P^o_1$ & 37939 & 39377 & 39427
\end{tabular}
\end{ruledtabular}
\end{table}
\subsection{Blackbody radiation}
It is well known that BBR-related clock frequency shifts arise from
perturbations of atomic energy levels by weakly oscillating thermal
radiation. For the $^1\!S_0 \rightarrow \, ^3\!P_0^o$ clock
transition, both atomic states involved are perturbed. Thus, the net
BBR effect is the difference of the individual BBR shifts for the
two states, $\delta\omega_\mathrm{BBR} = \delta
E_\mathrm{BBR}(^3\!P_0^o) - \delta E_\mathrm{BBR}(^1\!S_0)$. The
expression for the $\delta E_\mathrm{BBR}(g)$ of a $g$ state can be
given by the following formula~\cite{PorDer06BBR}
\begin{eqnarray}
\delta E_{\rm BBR}(g) = -\frac{2}{15} (\alpha \pi)^3 T^4
\alpha_{E_g}(0) \, [1 + \eta] \, ,  \label{Eq:delEg}
\end{eqnarray}
where $\alpha \approx 1/137$ is the fine structure constant, $T$ is
the characteristic temperature of the BBR environment,
$\alpha_{E_g}(0)$ is the electric-dipole static polarizability of
the $g$ state, and $\eta$ represents a ``dynamic'' fractional
correction to the total shift. As was shown in
Ref.~\cite{PorDer06BBR}, $\eta$ is negligible for the $^1\!S_0$
state but contributes to $\delta E_{\rm BBR}$ of the $^3\!P^o_0$
state at the 2.7\% level. This is primarily due to the fact that the
$^3\!P^o_0$ state has a transition frequency to a nearby $5s4d \,\,
^3D_1$ state in the infrared, as shown in Fig.~\ref{Fig:levels}.

The electric dipole static polarizabilities of the $5s^2
\,\,^1\!S_0$ and $5s5p \,\,^3\!P_0^o$ states were calculated in
Ref.~\cite{PorDer06BBR} using the same method, but a different basis
set, as presented in this work. The corresponding relative frequency
shift of the clock transition was determined to be $|\delta
\nu_{BBR} /\nu_0| = 55.0(7)\times 10^{-16}$.  The shift uncertainty
of $\Delta[\delta\nu_{BBR} /\nu_0]=0.7\times10^{-16}$ results
directly from the 1\% uncertainty attained in
Ref.~\cite{PorDer06BBR} for the static polarizability $\Delta
[\alpha_{^3\!P^o_0}(0)]$ of the $5s5p \,\,^3\!P_0^o$ state. The
small size of $\eta$ ensured that no additional uncertainty from the
dynamical correction contributed at this level.

An equally important source of uncertainty in the actual BBR shift
is the knowledge and control of the blackbody environment at room
temperature, $T$. From Eq.~(\ref{Eq:delEg}), the shift uncertainty
$\Delta[\delta E_{BBR}(g)]$ originating from the uncertainty in the
BBR environment $\Delta T$ is
\begin{eqnarray}
\Delta[\delta E_{\rm BBR}(g)] = -\frac{8}{15} (\alpha \pi)^3 T^3
\alpha_{E_g}(0) \, [1 + \eta] \Delta T \, ,  \label{Eq:delEg2}
\end{eqnarray}
At room temperature, measurement of the BBR environment at the
uncertainty level of $\Delta T$ = 1 K leads to a fractional
frequency shift uncertainty of $7.5 \times 10^{-17}$
~\cite{LudZelCam08}. The combination of this uncertainty in
quadrature with that resulting from $\Delta [\alpha_{^3\!P^o_0}(0)]$
yields a $1 \times 10^{-16}$ total BBR uncertainty, which currently
limits the accuracy of the Sr optical clock.

To further improve the Sr accuracy, the total BBR uncertainty must
be reduced. This requires solving two main problems: i) measuring
and controlling the blackbody environment to much better than
$\Delta T$ = 1 K; ii) determining the differential static
polarizability to better than 1\% uncertainty.  The first of these
requires additional care and design in the experimental apparatus.
While the temporal stability of the BBR temperature is more or less
straightforward to achieve, the experimental difficulty originates
from achieving spatial homogeneity of the BBR temperature over
various functional areas of the vacuum chamber housing the atomic
sample. Typically, a large fraction of the $4 \pi$ sterradians of
solid angle around the atoms consists of glass viewports to
accommodate the optical access needed for various atomic
manipulations (laser cooling and trapping, loading into an optical
lattice, state preparation, etc.\ ).  These areas are more difficult
to precisely temperature stabilize than the remainder of the solid
angle typically composed of metallic vacuum chamber. Experimentally,
we observe that different parts of the Sr vacuum apparatus at JILA
can vary by as much as 1 K.  Furthermore, at the highest clock
accuracy level, it is important to account for the effect of the
transmissivity of glass viewports for visible and infrared radiation
from the ambient room on the blackbody environment seen by the
atoms.  One approach to reducing the uncertainty of the BBR shift is
to surround the atomic sample in a cryogenically-cooled shield
~\cite{OskDid06}. Doing so reduces both the magnitude and thus the
uncertainty of the BBR shift.  Another approach is to enclose the
atoms in a chamber closely resembling the blackbody cavities used
for thermal radiation metrological standards ~\cite{TeJes03}.  For
example, the optically-confined atoms can be transported in a moving
lattice from a main chamber to a smaller, blackbody cavity
~\cite{LudZelCam08}. By careful temperature control of this small
cavity made of highly-thermally-conductive material, excellent
temperature homogeneity can be maintained.  The very limited optical
access (for lattice laser and clock probes) enables the effective
emissivity of the cavity interior to be very close to unity.  To
reach the $10^{-17}$ clock uncertainty, the BBR environment must be
known at the part per thousand level at room temperature,
corresponding to a BBR temperature accuracy at the 100 mK level.

The differential polarizability must also be carefully determined to
higher accuracy.  In the case of cesium, a well-controlled d.c.
electric field has been used to induce a clock shift and determine
the differential static polarizability ~\cite{SimLau98}.  As well,
some atom interferometric techniques may hold promise for directly
measuring the differential polarizability at better than the 1\%
level ~\cite{CroSch07}. Here we address the improved determination
of the differential polarizability based on atomic structure
measurements. The uncertainty of the differential static
polarizability is determined by the uncertainties in the
polarizabilities of the two clock states. The static polarizability
of the ground state $\alpha_{^1\!S_0}(0)$ = 197.2(2)
a.u.~\cite{PorDer06BBR} is known with 0.1\% accuracy. Consequently,
the task at hand is to determine the static polarizability of the
$5s5p \,\,^3\!P_0^o$ state with a similar level of accuracy. This is
a key step towards Sr lattice clock operation at the $10^{-17}$
uncertainty level. We now discuss this problem in detail and present
a possible solution in the following sections.

\subsection{Calculation of electric dipole a.c. polarizabilities of
the $5s^2 \,\, {}^1\!S_0$ and $5s5p \,\, {}^3\!P^o_0$ states}

Using the wave functions of the low-lying states obtained as a
result of solving the multiparticle Schr\"odinger equation, we are
able to calculate a.c. polarizabilities of the $5s^2 \,\,^1\!S_0$
and the $5s5p \,\,^3\!P_J^o$ states. As one check of the quality of
our calculations, we can find the magic wavelengths: $\lambda_0$ at
which $\alpha_{^1\!S_0} (\lambda_0) = \alpha_{^3\!P_0^o}
(\lambda_0)$ and $\lambda_1$ at which $\alpha_{^1\!S_0} (\lambda_1)
= \alpha_{^3\!P_1^o} (\lambda_1)$ and compare these values against
the experimental results. In recent works these magic wavelengths
were determined with high precision to be $\lambda_0 (^1\!S_0 - \,
^3\!P^o_0) = 813.42735(40)$ nm~\cite{LudZelCam08} and $\lambda_1
(^1\!S_0 - \, ^3\!P^o_1(m_J=\pm1)) = 914(1)$ nm for linear
polarization ~\cite{IdoKat03}. Furthermore, our calculation can also
be checked against a recent measurement of the a.c. Stark shift
associated with the probe of the ($^1\!S_0 \rightarrow \,
^3\!P^o_0$) clock transition itself~\cite{LudZelCam08}.

We start with a brief description of the method used to calculate
the electric dipole polarizabilities. The equation for the a.c.
electric dipole polarizability of the $g$ state can be written in
the following form,
\begin{eqnarray}
\alpha_{E_g} ( \omega ) &=& 2\, \sum_k \frac { \left( E_k-E_g \right)
|\langle g |d_0| k \rangle|^2 }
      { \left( E_k-E_g \right)^2 - \omega^2 } \nonumber \\
 &=&
\sum_k \frac {|\langle g |d_0| k \rangle|^2 } {E_k - (E_g + \omega)}
+
 \sum_k \frac {|\langle g |d_0| k \rangle|^2 } {E_k - (E_g - \omega)} \nonumber \\
&\equiv& \frac{1}{2} \left\{ \alpha_{E_g + \omega}(0) +
                    \alpha_{E_g - \omega}(0) \right\} .
\label{Eqn_alpha}
\end{eqnarray}
The two terms in the bottom-line of Eq.~(\ref{Eqn_alpha}) can be
viewed as the static polarizabilities of the $g$ state calculated
for the shifted energy levels of $E_g+\omega$ and $E_g-\omega$,
respectively. Thus, our task is reduced to computation of these two
static polarizabilities.

Following Refs.~\cite{DerJohSaf99,PorDer03} we decompose an a.c.
polarizability into two parts,
\begin{equation}
\alpha(\omega) = \alpha^v(\omega) + \alpha^c(\omega).
\end{equation}
The first term describes excitations of the valence electrons. The
second term characterizes excitations of core electrons and includes
a small counter term related to excitations of core electrons to
occupied valence state. The core polarizability $\alpha^c$ was
calculated at $\omega$=0 to be $\alpha^c(0) = 5.4$
a.u.~\cite{PorDer06BBR}.  Since $\alpha^c$ contributes to the total
polarizability only at the level of a few percent and its dependence
on frequency is very weak, the value of 5.4 a.u. can also be used
for calculations of the total a.c. polarizabilities. This
approximation of a constant core polarizability over the relevant
frequency range introduces an additional uncertainty of  $<$0.1$\%$
to the total $^3\!P^o_0$ polarizability.

It is worth mentioning that the core is the same for the $5s^2
\,\,^1\!S_0$ and the $5s5p \,\,^3\!P_J^o$ states. For this reason
$\alpha^c_{^1\!S_0}(\omega) \approx \alpha^c_{^3\!P_J^o}(\omega)$
and we arrive at the following expression,
\begin{equation}
\alpha_{^1\!S_0}(\omega) - \alpha_{^3\!P_J^o}(\omega) \approx
\alpha^v_{^1\!S_0}(\omega) - \alpha^v_{^3\!P_J^o}(\omega).
\end{equation}

The method of calculation for the dynamic {\em valence}
polarizabilities $\alpha^v(\omega)$ is described elsewhere (see,
e.g.,~\cite{KozPorFla96,PorDer03}). Here we only briefly
recapitulate its main features. These polarizabilities are computed
with the Sternheimer~\cite{Ste50} or Dalgarno-Lewis \cite{DalLew55}
method implemented in the CI+MBPT framework. (Here we denote
$\Sigma$ and RPA corrections as the many-body perturbation theory
(MBPT) corrections.) Given the $g$ state wave function and energy
$E_g$, we find intermediate-state wave functions $\delta \psi_{\pm}$
from an inhomogeneous equation,
\begin{eqnarray}
|\delta \psi_{\pm} \rangle & = & \frac{1}{H_{\rm eff} - (E_g \pm \omega)}\,
 \sum_k | k \rangle \langle k | d_0 | g \rangle \nonumber \\
&=&  \frac{1}{H_{\rm eff}-(E_g \pm \omega)} \, d_0 | g \rangle .
\label{delpsi}
\end{eqnarray}
Using Eq.~(\ref{Eqn_alpha}) and $\delta \psi_{\pm}$ introduced
above, we obtain
\begin{equation}
\alpha^v (\omega ) = \frac{1}{2} \left( \langle g |d_0| \delta
\psi_+ \rangle + \langle g |d_0| \delta \psi_- \rangle \right) \, ,
\label{alpha2}
\end{equation}
where superscript $v$ emphasizes that only excitations of the
valence electrons are included in the intermediate-state wave
functions $\delta \psi_{\pm}$ due to the presence of $H_{\rm eff}$.

\begin{table}
\caption{Calculated polarizabilities at a few selected optical
wavelengths. Wavelengths $\lambda$ are in nm, the frequencies
$\omega$ are in a.u. and the electric dipole a.c. polarizabilities
of the $5s^2 \,\, ^1\!S_0$ and the $5s5p \,\, ^3\!P^o_{0,1}$ states
are in a.u. The polarizability of the $^3\!P^o_1$ state is
calculated for the projection $|m_J|=1$ and linearly polarized
light.} \label{Polariz}
\begin{ruledtabular}
\begin{tabular}{ccccc}
   \multicolumn{1}{c}{$\lambda$}
 & \multicolumn{1}{c}{$\omega$}
 & \multicolumn{1}{c}{$\alpha(5s^2 \,\, {}^1\!S_0)$}
 & \multicolumn{1}{c}{$\alpha(5s5p \,\, {}^3\!P_0^o)$}
 & \multicolumn{1}{c}{$\alpha(5s5p \,\, {}^3\!P_1^o)$} \\
\hline
        &  0.0000  &  197.2  &  457.0  &  498.8 \\
 698.4  &  0.0652  &  351.8  &  909.2  &        \\
 805.0  &  0.0566  &  288.9  &  289.3  &        \\
 813.4  &  0.0560  &  286.0  &  280.5  &        \\
\hline
 902.2  &  0.0505  &  263.5  &         &  263.4 \\
 914.0  &  0.0499  &  261.2  &         &  256.1 \\
\end{tabular}
\end{ruledtabular}
\end{table}

In Table~\ref{Polariz} we present the values of the static
polarizabilities and the a.c. polarizabilities of the $5s^2
\,\,^1\!S_0$ and the $5s5p \,\,^3\!P_{0,1}^o$ states computed for
different values of $\lambda$ using the CI+MBPT approach. As a first
step we solved an inhomogeneous equation and found the valence parts
of the polarizabilities. Then the values of the polarizabilities of
the ground state were corrected as follows. We used the fact that
the intermediate $5s5p \,\,^1\!P_1^o$ state contributes to this
polarizability at the level of 97\%. Knowing the experimental energy
difference ($E_{^1\!P_1^o} - E_{^1\!S_0}$) and the matrix element
$|\langle 5s^2 \,\,^1\!S_0 ||d|| 5s5p \,\,^1\!P_1^o \rangle|$ =
5.249(2) a.u. extracted from the precise measurement of the lifetime
of the $5s5p \,\,^1\!P_1^o$ state~\cite{YasKisTak06}, we replaced
the theoretical contribution of the $5s5p \,\,^1\!P_1^o$ state to
the ground-state polarizability by the experimental value. Finally,
we added $\alpha^c$ term to the valence parts, arriving at the
values listed in Table~\ref{Polariz}.

Starting from the 0.05\% uncertainty of the $|\langle 5s^2
\,\,^1\!S_0 ||d|| 5s5p \,\,^1\!P_1^o \rangle|$ matrix element, we
estimated the uncertainty of the a.c. polarizability of the ground
state at the level of 0.1\%. In particular, for the static
polarizability, we obtained $\alpha_{^1\!S_0}(0)$ = 197.2 a.u., in
perfect agreement with the result obtained in
Ref.~\cite{PorDer06BBR} using a different basis set.

\begin{figure}[h]
\begin{center}
\includegraphics*[scale=0.35]{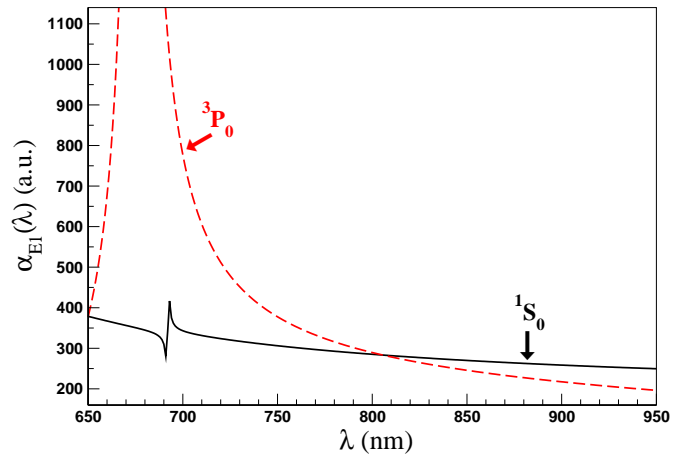}
\end{center}
\caption{ (Color online) Electric dipole a.c.\ polarizabilities for
$5s^2 \,\, ^1\!S_0$ (solid line) and $5s5p \,\,^3\!P_0$ (dashed
line) states of Sr. The polarizabilities are shown as a function of
optical wavelength $\protect\lambda$. } \label{Fig:dynPol}
\end{figure}

Experiments~\cite{LudZelCam08,IdoKat03} have determined the magic
wavelengths for the $^1\!S_0 - \, ^3\!P_0^o$ and $^1\!S_0 - \,
^3\!P_1^o(m_J=\pm1)$ transitions to be 813.4 nm and 914 nm,
respectively. As is seen from Table~\ref{Polariz}, the calculations
carried out in this work give the values of 805 nm and 902 nm for
these magic wavelengths, respectively. Thus, the agreement between
theoretical and experimental results is at the level of 1\%. The
behavior of the a.c. polarizabilities of the $^1\!S_0$ and the
$^3\!P_0^o$ states in the wavelength range from 650 nm to 950 nm is
illustrated in Fig.~\ref{Fig:dynPol}. A large peak at 679 nm for the
$^3\!P_0^o$ state arises from the contribution of the $5s6s \,\,
^3\!S_1$ state, while a small peak in the vicinity of 690 nm for the
$^1\!S_0$ a.c. polarizability is due to the contribution of the
$5s5p \,\, ^3\!P^o_1$ state. Experimentally, the differential a.c.
polarizabilities in the form of the clock frequency shift induced by
the clock probe laser itself is known. With a probe laser intensity
of $~$20 mW/cm$^2$, the fractional frequency shift was measured
-1.5(0.4)$\times$ 10$^{-15}$~\cite{LudZelCam08}. Assuming the same
probe laser intensity and the values of the polarizabilities for the
$^1\!S_0$ and $^3\!P^o_1$ states obtained at 698.4 nm (see
Table~\ref{Polariz}), the calculated fractional shift is -1.2
$\times$ 10$^{-15}$, in a good agreement with experiment.

It is worth noting that knowing the precise experimental values of
the magic wavelengths and using the fact that
$\alpha_{^3\!P_0^o}(813.4 \,\, {\rm nm})$ = $\alpha_{^1\!S_0}(813.4
\,\, {\rm nm})$ and $\alpha_{^3\!P_1^o,|m_J|=1}(914\,\, {\rm nm})$ =
$\alpha_{^1\!S_0}(914\,\, {\rm nm})$, we can refine our calculation
and predict with high accuracy the values of the a.c.
polarizabilities of the $^3\!P_0^o$ and $^3\!P_1^o$ states at these
wavelengths. We obtain $\alpha_{^3\!P_0^o}(813.4 \,\, {\rm nm})$ =
286.0(3) a.u. and $\alpha_{^3\!P_1^o,|m_J|=1}(914\,\, {\rm nm})$ =
261.2(3) a.u., matching the polarizabilities of the $^1\!S_0$ state
at these two wavelengths.

\begin{table*}
\caption{Individual contributions (I.C.) from six low-lying
intermediate states to the valence parts of the static
polarizability $\alpha^v_{^3\!P^o_0}(0) = 451.5$ a.u. and to the
a.c. polarizabilities (in a.u.) at the wavelengths $\lambda$ = 813.4
nm and $\lambda$ = 698.4 nm. $\alpha^v_{^3\!P^o_0}$(813.4 nm) =
280.6 a.u. and $\alpha^v_{^3\!P^o_0}$ (698.4 nm) = 903.8 a.u.. $D
\equiv |\langle 5s5p \,\, ^3\!P^o_0 ||d|| n \rangle|$ is the reduced
matrix element of the electric dipole operator ${\bf d}$. The row
``Total'' gives the sum of the contributions for each column.}
\label{Ind_contrib}
\begin{ruledtabular}
\begin{tabular}{lccccccc}
&& \multicolumn{2}{c}{I.C. to $\alpha^v_{^3\!P^o_0}(0)$}
 & \multicolumn{2}{c}{I.C. to $\alpha^v_{^3\!P^o_0}(813.4\, {\rm nm})$}
 & \multicolumn{2}{c}{I.C. to $\alpha^v_{^3\!P^o_0}(698.4\, {\rm nm})$}  \\
   \multicolumn{1}{c}{$|n\rangle$}
 & \multicolumn{1}{c}{ $D$ (a.u.)}
 & \multicolumn{1}{c}{a.u.}  & \multicolumn{1}{c}{\%}
 & \multicolumn{1}{c}{a.u.}  & \multicolumn{1}{c}{\%}
 & \multicolumn{1}{c}{a.u.}  & \multicolumn{1}{c}{\%} \\
\hline
$5s4d \,\,^3D\!_1$ &  2.74 &  286.1  & 63.4 &  -31.0  &-11.3  & -22.2  & -2.5\\
$5s6s \,\,^3S\!_1$ &  1.96 &   38.3  &  8.5 &  126.4  & 45.9  & 705.8  & 78.1\\
$5s5d \,\,^3D\!_1$ &  2.50 &   44.2  &  9.8 &   68.3  & 24.8  &  84.8  &  9.4\\
$5p^2 \,\,^3P\!_1$ &  2.56 &   45.3  & 10.0 &   68.7  & 25.0  &  84.1  &  9.3\\
$5s7s \,\,^3S\!_1$ &  0.52 &    1.7  &  0.4 &    2.4  &  0.9  &   2.8  &  0.3\\
$5s6d \,\,^3D\!_1$ &  1.13 &    7.4  &  1.6 &    9.6  &  3.5  &  10.8  &  1.2\\
\hline
Total              &       &  423.4  & 93.8 &  245.8  & 89.3  & 868.4  & 96.1\\
\end{tabular}
\end{ruledtabular}
\end{table*}


In Table~\ref{Ind_contrib} we present the values of the individual
contributions from six low-lying even-parity intermediate states to
the valence parts of the static polarizability and the a.c.
polarizabilities at $\lambda$ = 813.4 nm and $\lambda$ = 698.4 nm
for the $^3\!P^o_0$ state. As is seen from Table~\ref{Ind_contrib},
for the $^3\!P^o_0$ static polarizability, four even-parity states
($5s4d \,\,^3\!D_1$, $5s6s \,\,^3\!S_1$, $5s5d \,\,^3\!D_1$, and
$5p^2 \,\,^3\!P_1$) contribute at the level of 92\%. If the sum of these four
contributions are determined experimentally with an accuracy
$\sim$0.1\% and the contributions of all the rest of the discrete
and continuum states are known at the level of 1-2\% from
calculations, the overall polarizability of the $5s5p \,\,
^3\!P^o_0$ state can be determined with the accuracy $\sim$0.1\%.

Experimental determination of the four dominant contributions may be
accomplished directly from lifetime or transition rate measurements.
However, the lifetime data should be accompanied by high accuracy
branching ratio measurements. Alternatively, we could measure single
channel decay directly to the $5s5p \,\,^3\!P^o_0$ state. In
addition, it is possible to constrain the four contributions using
other spectroscopic data such as the magic wavelength $\lambda_0$
and the light shift at 698 nm, which will naturally be measured and
confirmed by a number of future clock experiments.  At both
wavelength regions the same four states dominate the polarizability
as in the static case. The general strategy would be to use
spectroscopic data to constrain the most dominant contribution for
that specific case.

Since the $5s4d \,\,^3\!D_1$ state dominates the four critical
contributions to the static polarizability, and has less
contribution to the dynamic polarizability at wavelengths of
interest (see Table~\ref{Ind_contrib}), it would thus be maximally
beneficial to measure this contribution directly with an oscillator
strength measurement between $5s5p \,\,^3\!P^o_0$ and $5s4d
\,\,^3\!D_1$. Doing so avoids upscaling of the uncertainty via error
propagation. The high accuracy experimental measurement for the
$5s5d \,\,^3\!D_1$ state~\cite{Andra75} can be further improved by
monitoring its decay to $5s5p \,\,^3\!P^o_0$ directly. Then,
combining the measurements of $\lambda_0$ and the light shift at 698
nm would allow us to constrain contributions from both $5s6s
\,\,^3\!S_1$ and $5p^2 \,\,^3\!P_1$, permitting constraining
$\alpha_{^3\!P^o_0}(0)$ at the level of $\alpha_{^1\!S_0}(0)$.

The magic wavelength, $\lambda_1$, could in principle also aid the
constraint. However, additional sensitivities of the $5s5p
\,^3\!P_1^o$ polarizability from $J$ = 0 and $J$ = 2 even-parity
states are large and complex. Furthermore, the vector nature of this
state requires careful experimental control of light polarization.

In the following section we present the results of the theoretical
calculation of transition rates and oscillator strengths most
relevant to the $^3\!P^o_0$ and $^1\!S_0$ polarizabilities and
compare them to existing experimental and theoretical data in the
literature.

\subsection{Transition rates and oscillator strengths.}
\label{Tr_rates}
The transition rate ($W$) and the oscillator strength ($f$) for a
transition from an initial state $|\gamma' J' L' S'\rangle$ to a
final state $\langle \gamma J L S|$ can be represented as (see,
e.g.,~\cite{Sob79})
\begin{eqnarray}
&&W(\gamma' J' L' S' \rightarrow \gamma J L S) = \nonumber \\
&& \frac{4\,(\omega \alpha)^3}{3} \frac{1}{2J'+1}
 |\langle \gamma J L S ||d|| \gamma' J' L' S'\rangle|^2,
\label{W_JLS}
\end{eqnarray}
\begin{eqnarray}
&& f(\gamma' J' L' S' \rightarrow \gamma J L S) =  \nonumber \\
&& -\frac{2 \,\omega}{3}
\, \frac{1}{2J'+1} |\langle \gamma J L S ||d|| \gamma' J' L'
S'\rangle|^2,
\label{f_JLS}
\end{eqnarray}
where $\gamma$ denotes all quantum numbers other than $J$, $L$, and
$S$. $\omega = E_{\gamma' J'} - E_{\gamma J}$ is the transition
frequency from the initial state to the final state. With this
definition the oscillator strength is positive for absorption and
negative for emission.

\begin{table*}
\caption{Transition rates $W_{n \rightarrow 0}$ and oscillator
strengths $f_{0 \rightarrow n}$ for relevant energy levels in Sr.
The results are compared with other available experimental and
theoretical data.} \label{TranRate_OscSt}
\begin{ruledtabular}
\begin{tabular}{lcccdd}
&& \multicolumn{2}{c}{$W_{n \rightarrow 0}(\times \, 10^6\,s^{-1})$}
 & \multicolumn{2}{c}{$f_{0 \rightarrow n}$} \\
   \multicolumn{1}{c}{$\langle 0|$}
 & \multicolumn{1}{c}{$|n \rangle$}
 & \multicolumn{1}{c}{This work} & \multicolumn{1}{c}{Other data}
 & \multicolumn{1}{c}{This work} & \multicolumn{1}{c}{Other data \footnotemark[1]} \\
\hline $5s^2 \,\,^1\!S_0$ & $5s5p \,\,^1\!P_1^o$
                   & 186.0\footnotemark[2]&           190.01(14) \footnotemark[3]
                                                                  & 1.82\footnotemark[2]    & 1.92(6)    \\
                   &                      &         & 191.6(1.1) \footnotemark[4] &         &            \\
                   &                      &         & 215        \footnotemark[5] &         &            \\
                   & $5s6p \,\,^1\!P_1^o$ &  1.49   &   1.87(26) \footnotemark[1] & 0.0058  & 0.0072(10) \\
                   &                      &         &   3.79     \footnotemark[5] &         &            \\
                   & $5s7p \,\,^1\!P_1^o$ &  5.13   &   5.32(61) \footnotemark[1] & 0.15    & 0.16(2)    \\
                   &                      &         &   3.19     \footnotemark[5] &         &            \\
\hline
$5s5p\,\,^3\!P_0^o$& $5s4d \,\,^3\!D_1$   &  0.29   &            & 0.088   &            \\
                   & $5s6s \,\,^3\!S_1$   &  8.39   &   7.32     \footnotemark[5] & 0.173   &            \\
                   & $5s5d \,\,^3\!D_1$   & 38.1    &   30.7     \footnotemark[5] & 0.395   &            \\
                   & $5p^2 \,\,^3\!P_1$   & 41.3    &                             & 0.418   &            \\
                   & $5s7s \,\,^3\!S_1$   &  2.28   &    1.80    \footnotemark[5] & 0.019   &            \\
                   & $5s6d \,\,^3\!D_1$   & 14.3    &                             & 0.099   &            \\
\end{tabular}
\end{ruledtabular}
\small\footnotemark[1]{Parkinson {\it et al.}~\cite{ParReeTom76}(exp.).} \\
\small\footnotemark[2]{These values are presented only for
comparison. In all calculations performed in this work involving
$\langle ^1\!S_0 ||d|| ^1\!P_1^o \rangle$, we used the value of this
ME obtained from Ref.~\cite{YasKisTak06}.}\\
\small\footnotemark[3]{Yasuda {\it et al.}~\cite{YasKisTak06}(exp.).} \\
\small\footnotemark[4]{Nagel {\it et al.}~\cite{NagMicSae05}(exp.).} \\
\small\footnotemark[5]{These numbers were obtained from the values
given in~Werij {\it et al.}~\cite{WerGreThe92} with use of
Eqs.~(\ref{W_JLS}) and (\ref{MatEl})
(see Subsection~\ref{Tr_rates} for details).} \\
\end{table*}

Using Eqs.~(\ref{W_JLS}) and (\ref{f_JLS}) and knowing $E1$
transition amplitudes between different states, we were able to
calculate rates and oscillator strengths for the transitions
involving the $5s^2 \,\, ^1\!S_0$ and the $5s5p \,\, ^3\!P^o_0$
states. In Table~\ref{TranRate_OscSt} we list the transition rates
and oscillator strengths for the strongest transitions from the
mentioned states. Note that the transition rates $W_{n \rightarrow
^1\!S_0, ^3\!P^o_0}$ and the oscillator strengths $f_{^1\!S_0,
^3\!P^o_0 \rightarrow n}$ were calculated with use of the {\it
theoretical} energy levels. Where available we compare our results
with other experimental and theoretical values. As is seen from
Table~\ref{TranRate_OscSt}, there is a reasonable agreement between
the results of this work and other data.

In certain cases we used for comparison the non-relativistic values
of transition rates given in the literature. In the $LS$ coupling
approximation, there is a simple relation between relativistic and
non-relativistic reduced matrix elements (MEs) of the operator ${\bf
d}$. Since the operator ${\bf d}$ commutes with ${\bf S}$ we
obtain~\cite{Sob79},
\begin{eqnarray}
&&\langle \gamma J L S ||d|| \gamma' J' L' S'\rangle =
 \delta_{SS'} \sqrt{(2J+1)(2J'+1)}   \nonumber \\
& \times & (-1)^{S+L+J'+1} \, \left\{
\begin{array}{ccc}
L  & J  & S  \\
J' & L' & 1
\end{array}
\right\} \,
\langle \gamma L S ||d|| \gamma' L' S\rangle.
\label{MatEl}
\end{eqnarray}
Knowing a non-relativistic transition rate we were able to determine
the corresponding ME of the electric dipole operator $\langle \gamma
L S ||d|| \gamma' L' S\rangle$. If the $LS$ coupling approximation
is valid, using Eq.~(\ref{MatEl}), the non-relativistic ME can be
related to the relativistic ME $\langle \gamma J L S ||d|| \gamma'
J' L' S'\rangle$. We also need to account for the fact that in a
non-relativistic case the transition frequency $\omega_{LS}$ between
two states is given by expression $\omega_{LS} =
\overline{E_{\gamma' J' L' S'}} - \overline{E_{\gamma J L S}}$,
where $\overline{E_{\gamma J L S}}$ is the center of gravity of the
respective multiplet. For this reason, in general, $\omega_{LS}$ can
slightly differ from $\omega= E_{\gamma' J'} - E_{\gamma J}$.
Finally, using Eq.~(\ref{W_JLS}) we can find the relativistic
transition rate. We used this approach to compare the relativistic
transition rates obtained in this work with the non-relativistic
values presented in~\cite{WerGreThe92}.

If the $LS$ coupling breaks down, Eq.~(\ref{MatEl}) is no longer
valid. In this case, to compare the relativistic transition rates
given by Eq.(\ref{W_JLS}) with the non-relativistic transition rates
$W(\gamma' L' S' \rightarrow \gamma L S)$, one should use a more
general relation~\cite{Sob79}:
\begin{eqnarray}
&& W(\gamma' L' S' \rightarrow \gamma L S) =
\frac{1}{(2L'+1)(2S'+1)} \times  \nonumber \\
&&\sum_{JJ'} (2J'+1)
W(\gamma' J' L' S' \rightarrow \gamma J L S).
\label{WW}
\end{eqnarray}
In the right-hand side of this equation the summation goes over all possible
values of $J'$ and $J$. Consequently, we need to find all
transitions rates (permitted by selection rules) from the fine
structure levels of one multiplet to the fine structure levels of
another multiplet.

To provide a straightforward comparison of the calculation here with
experimental data, lifetimes of the four states which dominate the
$^3\!P^o_0$ polarizability contributions have been evaluated.  For
these four states, decay to the $5s5p \,\, ^3\!P^o_J$ states is the
only significant radiative decay channel so the lifetimes can
provide direct information on the relevant matrix elements. A number
of lifetime and transition rate measurements are available for
comparison \cite{WerGreThe92,Fakuda,
MilYouCoo92,BorPenRed87,JonLevPer84,Brinkmann69,Havey77,Andra75,Osherovich79,Gornik77},
however in many instances with limited accuracy.  Table~\ref{EXP}
summarizes the results where the $5s4d \,\, ^3\!D_1$, $5s6s \,\,
^3\!S_1$, $5s5d \,\, ^3\!D_1$, and $5p^2 \,\, ^3\!P_1$ lifetime
calculations are compared to available measurements.  In some cases,
the measured lifetimes were reported for a particular $J$ value in
the excited state multiplet, and in others only a mean lifetime for
the entire multiplet was given, leading to complications in the
analysis.

In the case of the $5s4d \,\, ^3\!D$ state we also evaluated the
total transition rate for the multiplet since to our knowledge the
only lifetime measurements for the $5s4d \,\, ^3\!D$ levels were
performed on the entire multiplet. In the framework described above
we found the $W(5s4d \,\, ^3\!D_{J\,'} \rightarrow \, 5s5p\,\,
^3\!P_J^o)$ transition rates for all possible $J$ and $J'$, and
using Eq.(\ref{WW}), we obtained,
$$
\frac{1}{15} \sum_{JJ'} (2J\,'+1) W(5s4d \,\, ^3\!D_{J\,'}
\rightarrow \, 5s5p\,\, ^3\!P_J^o) = 0.41 \times 10^6 {\rm s}^{-1}
$$
in agreement with the experimental value $W(5s4d \,\, ^3\!D
\rightarrow \, 5s5p\,\, ^3\!P^o)
 = 0.345(24) \times 10^6$ s$^{-1}$~\cite{MilYouCoo92}.

The lifetimes of the $5s5d \,\, ^3\!D_J$ states have been measured
with high accuracy (0.1\%) in Ref.~\cite{Andra75} and are in good
agreement with other measurements as well as our calculated values.
Ref.~\cite{Andra75} also reported a 1\% measurement of the
$5p^2\,\,^3\!P_2$ lifetime, while other direct measurements of this
multiplet yield consistent $^3\!P_1$ lifetimes having accuracies at
the 15-20 \% level. Relative to $5s5d \,\, ^3\!D_J$ and
$5p^2\,\,^3\!P_J$, the $5s6s \,\, ^3\!S_1$ experimental data has
larger scatters between different measurements. Notably all of these
measured lifetime values agree well with our calculations.

Given the results in Table~\ref{EXP} and Table~\ref{Ind_contrib},
the $5s4d \,\, ^3\!D_1$ state should have the highest measurement
priority as it dominates the $^3\!P_0^o$ static polarizability. It
also has a large disagreement between the experiments~
\cite{MilYouCoo92} and~\cite{BorPenRed87}, meriting further
experimental investigations. The next priority goes to the $5s6s
\,\, ^3\!S_1$ state due to the scatter in existing data and its
large contribution to the a.c. polarizability of $^3\!P_0^o$.
Perhaps a good strategy is to measure its decay directly to
individual $5s5p \,\, ^3\!P^o_J$ states. Confirmation of the high
accuracy result of Ref.~\cite{Andra75} for the $5s5d \,\, ^3\!D_1$
state and improvement upon the $5p^2 \,\, ^3\!P_1$ result listed in
Table~\ref{EXP}, or alternatively, the use of the measured magic
wavelength and clock laser light shift, can then be sufficient to
determine the $^3\!P_0^o$ polarizability at the 0.1\% level.

\begin{table}
\caption{Experimental lifetime data for the low-lying even-parity
states. Theory values calculated in this work are provided for
comparison. $\overline{^3\!D}$ denotes measurement of the $^3\!D$
manifold without definitive angular momenta.} \label{EXP}
\begin{ruledtabular}
\begin{tabular}{rcc}
   \multicolumn{1}{c}{Excited state}
 & \multicolumn{1}{c}{$\tau$ (ns)}
 &\multicolumn{1}{c}{This work}\\
\hline \\ $5s4d \,\,^3\!D_1$  &                                  & 2040  \\&&\\
         $ \overline{^3\!D}$  &   2900(200) \protect\footnotemark[1]     & 2400 \footnotemark[2] \\
                              &   4100(600) \footnotemark[3]     &       \\

\hline \\ $5s6s \,\,^3\!S_1$  &   15.0(8) \footnotemark[4]       &  14.1  \\
                              &   10.9(1.1) \footnotemark[5]     &        \\
                              &   12.9(7) \footnotemark[6]       &        \\

\hline \\$5s5d \,\,^3\!D_1$   &   16.49(10) \footnotemark[7]     & 16.9   \\&&\\
                 $ ^3\!D_2$   &   16.34(13) \footnotemark[7]     & 16.8   \\&&\\
                 $ ^3\!D_3$   &   16.29(24) \footnotemark[7]     &        \\&&\\
        $ \overline{^3\!D}$   &   17.1(8)   \footnotemark[8]     &        \\
                              &   16.0(6)   \footnotemark[4]     &        \\
                              &   16.7(1.0) \footnotemark[6]     &        \\

\hline \\ $5p^2 \,\,^3\!P_0$  &                                  &  7.8   \\ &&\\
                  $ ^3\!P_1$  &    8.3(0.4) \footnotemark[4]     &  7.6   \\
                              &    8.8(1.2) \footnotemark[5]     &        \\
                              &   10.2(2.4) \footnotemark[9]     &        \\ &&\\
                  $ ^3\!P_2$  &    7.89(05) \footnotemark[7]     &  7.9   \\
                              &    7.8(1.8)  \footnotemark[5]    &        \\
                              &    8.3(0.4)  \footnotemark[4]    &        \\ 
\end{tabular}
\end{ruledtabular}
\footnotetext[1]{Reference~\cite{MilYouCoo92}.}
\footnotetext[2]{This value is calculated using Eq.~(\ref{WW}).}
\footnotetext[3]{Reference~\cite{BorPenRed87}.}
\footnotetext[4]{Reference~\cite{JonLevPer84}.}
\footnotetext[5]{Reference~\cite{Brinkmann69}.}
\footnotetext[6]{Reference~\cite{Havey77}.}
\footnotetext[7]{Reference~\cite{Andra75}.}
\footnotetext[8]{Reference~\cite{Osherovich79}.}
\footnotetext[9]{Reference~\cite{Gornik77}.}
\end{table}
\section{Conclusion}
In this work we have carried out detailed calculations in response
to the goal of further improving the accuracy of the Sr atomic clock
to 1$\times$10$^{-17}$ and below. To focus on the outstanding
problem of BBR-related frequency shifts, we calculated a.c.
polarizabilities of the $^1\!S_0$ and $^3\!P_0^o$ clock states. We
verify our calculations with available experimental data. For
example, the theoretically calculated magic wavelengths for the
$^1\!S_0 - \,^3\!P_0^o$ and the $^1\!S_0 - \,^3\!P_1^o$ transitions
are in 1\% agreement with experiments. The agreement between theory
and experiment on the a.c. Stark shift of the clock transition
itself is also good. We have calculated individual contributions of
six lowest-lying even-parity states to the polarizability of the
$^3\!P_0^o$ state at $\omega=0$ and for the wavelengths $\lambda$ =
698.4 nm and 813.4 nm. We determined four even-parity states whose
total contribution to the static polarizability of the $^3\!P_0^o$
clock state is $\sim$ 90\%. Using the modern methods of atomic
calculations we can find the contribution of all the other discrete
and continuum states (constituting 10\%) to the $^3\!P_0^o$
polarizability at the level of 1-2\%. For this reason, if the
contributions of the four states identified here are experimentally
determined with 0.1\% accuracy, the same level of accuracy can be
obtained for the total polarizability of $^3\!P_0^o$. In the near
future we plan to undertake experimental measurements to determine
the oscillator strengths for the four identified states.
Measurements could include transition linewidths, power broadening
coefficients, direct lifetime determinations of individual $J$
levels, and improved determination of the clock laser a.c. Stark
shifts. These experimental measurements can be further combined with
the well-known value of $\lambda_0$. The experimentally determined
values will be used to refine the theory calculations presented here
to reach the goal of determining the polarizability of $^3\!P_0^o$
at 0.1\%.

\begin{acknowledgments}
We thank C. Greene and P. Lemonde for useful discussions and the
rest of JILA Sr researchers for their contributions to the
experimental work cited here. S.G.P. would like to thank JILA for
hospitality and in particular the JILA Visiting Fellows Program for
partial financial support. S.G.P. was also supported in part by the
Russian Foundation for Basic Research under Grants No.~07-02-00210-a
and No.~08-02-00460-a. We acknowledge funding support from NSF,
NIST, and DARPA.
\end{acknowledgments}


\begin{thebibliography}{30}
\expandafter\ifx\csname
natexlab\endcsname\relax\def\natexlab#1{#1}\fi
\expandafter\ifx\csname bibnamefont\endcsname\relax
  \def\bibnamefont#1{#1}\fi
\expandafter\ifx\csname bibfnamefont\endcsname\relax
  \def\bibfnamefont#1{#1}\fi
\expandafter\ifx\csname citenamefont\endcsname\relax
  \def\citenamefont#1{#1}\fi
\expandafter\ifx\csname url\endcsname\relax
  \def\url#1{\texttt{#1}}\fi
\expandafter\ifx\csname urlprefix\endcsname\relax\def\urlprefix{URL
}\fi \providecommand{\bibinfo}[2]{#2}
\providecommand{\eprint}[2][]{\url{#2}}



\bibitem{Katori03}
H. Katori {\it et al.}, Phys. Rev. Lett. {\bf91}, 173005 (2003).

\bibitem[{\citenamefont{{A. D. Ludlow \it et al.}}(2006)}]{ludlow06}
\bibinfo{author}{\bibnamefont{{A. D. Ludlow \it et al.}}},
  \bibinfo{journal}{Phys. Rev. Lett.} \textbf{\bibinfo{volume}{96}},
  \bibinfo{pages}{033003} (\bibinfo{year}{2006}).

\bibitem{Boyd2}
    M. M. Boyd {\it et al.}, Phys. Rev. Lett. {\bf 98}, 083002
    (2007).

\bibitem{Campbell}
G. K. Campbell {\it et al.}, arXiv:physics/0804.4509.

\bibitem[{\citenamefont{{R. {Le Targat} \it et al.}}(2006)}]{letargat06}
\bibinfo{author}{\bibnamefont{{R. {Le Targat} \it et al.}}},
  \bibinfo{journal}{Phys. Rev. Lett.} \textbf{\bibinfo{volume}{97}},
  \bibinfo{pages}{130801} (\bibinfo{year}{2006}).

\bibitem{Baillard}
X. Baillard {\it et al.}, Eur. Phys. J. D {\bf 48}, 11 (2008).

\bibitem[{\citenamefont{{M. Takamoto \it et al.}}(2006)}]{takamoto06}
\bibinfo{author}{\bibnamefont{{M. Takamoto \it et al.}}}, \bibinfo{journal}{J.
  Phys. Soc. Jpn.} \textbf{\bibinfo{volume}{75}}, \bibinfo{pages}{104302}
  (\bibinfo{year}{2006}).

\bibitem[{\citenamefont{Ludlow et~al.}(2008)\citenamefont{Ludlow, Zelevinsky,
  Campbell, Blatt, Boyd, {\rm de Miranda}, Martin, Thomsen, Foreman, Ye
  et~al.}}]{LudZelCam08}
\bibinfo{author}{\bibfnamefont{A.~D.} \bibnamefont{Ludlow}},
  \bibinfo{author}{\bibfnamefont{T.}~\bibnamefont{Zelevinsky}},
  \bibinfo{author}{\bibfnamefont{G.~K.} \bibnamefont{Campbell}},
  \bibinfo{author}{\bibfnamefont{S.}~\bibnamefont{Blatt}},
  \bibinfo{author}{\bibfnamefont{M.~M.} \bibnamefont{Boyd}},
  \bibinfo{author}{\bibfnamefont{M.~{\rm H. G}.} \bibnamefont{{\rm de
  Miranda}}}, \bibinfo{author}{\bibfnamefont{M.~J.} \bibnamefont{Martin}},
  \bibinfo{author}{\bibfnamefont{J.~W.} \bibnamefont{Thomsen}},
  \bibinfo{author}{\bibfnamefont{S.~M.} \bibnamefont{Foreman}},
  \bibinfo{author}{\bibfnamefont{J.}~\bibnamefont{Ye}}, \bibnamefont{et~al.},
  \bibinfo{journal}{Science} \textbf{\bibinfo{volume}{319}},
  \bibinfo{pages}{1805} (\bibinfo{year}{2008}).

\bibitem{NistYb2}
    N. Poli {\it et al.}, Phys. Rev. A {\bf 77}, 050501 (2008).

\bibitem[{\citenamefont{{J. Ye, H. J. Kimble, and H. Katori}}(2008{\natexlab{b}})}]{Ye08}
\bibinfo{author}{\bibnamefont{{J. Ye, H. J. Kimble, H. Katori}}},
  \bibinfo{journal}{Science} \textbf{\bibinfo{volume}{320}},
  \bibinfo{pages}{1734} (\bibinfo{year}{2008}{\natexlab{b}}).

\bibitem[{\citenamefont{{M. M. Boyd \it et al.}}(2006{\natexlab{b}})}]{boyd06}
\bibinfo{author}{\bibnamefont{{M. M. Boyd \it et al.}}},
  \bibinfo{journal}{Science} \textbf{\bibinfo{volume}{314}},
  \bibinfo{pages}{1430} (\bibinfo{year}{2006}{\natexlab{b}}).

\bibitem[{\citenamefont{Porsev and Derevianko}(2004)}]{PorDer04}
\bibinfo{author}{\bibfnamefont{S.}~\bibnamefont{Porsev}} \bibnamefont{and}
  \bibinfo{author}{\bibfnamefont{A.}~\bibnamefont{Derevianko}},
  \bibinfo{journal}{Phys. Rev. A} \textbf{\bibinfo{volume}{69}},
  \bibinfo{pages}{042506} (\bibinfo{year}{2004}).

\bibitem[{\citenamefont{{R. Santra \it et al.}}(2004{\natexlab{b}})}]{santra04}
\bibinfo{author}{\bibnamefont{{R. Santra \it et al.}}},
  \bibinfo{journal}{Phys. Rev. A} \textbf{\bibinfo{volume}{69}},
  \bibinfo{pages}{042510} (\bibinfo{year}{2004}{\natexlab{b}}).

\bibitem{BoydPRA}
    M. M. Boyd {\it et al.}, Phys. Rev. A {\bf 76}, 022510 (2007).

\bibitem[{\citenamefont{Porsev and Derevianko}(2006)}]{PorDer06BBR}
\bibinfo{author}{\bibfnamefont{S.~G.} \bibnamefont{Porsev}} \bibnamefont{and}
  \bibinfo{author}{\bibfnamefont{A.}~\bibnamefont{Derevianko}},
  \bibinfo{journal}{Phys. Rev. A} \textbf{\bibinfo{volume}{74}},
  \bibinfo{pages}{020502(R)} (\bibinfo{year}{2006}).

\bibitem[{\citenamefont{Yasuda et~al.}(2006)\citenamefont{Yasuda, Kishimoto,
  Takamoto, and Katori}}]{YasKisTak06}
\bibinfo{author}{\bibfnamefont{M.}~\bibnamefont{Yasuda}},
  \bibinfo{author}{\bibfnamefont{T.}~\bibnamefont{Kishimoto}},
  \bibinfo{author}{\bibfnamefont{M.}~\bibnamefont{Takamoto}}, \bibnamefont{and}
  \bibinfo{author}{\bibfnamefont{H.}~\bibnamefont{Katori}},
  \bibinfo{journal}{Phys. Rev. A} \textbf{\bibinfo{volume}{73}},
  \bibinfo{pages}{011403} (\bibinfo{year}{2006}).

\bibitem[{\citenamefont{Grant and Quiney}(1988)}]{GraQui88}
\bibinfo{author}{\bibfnamefont{I.~P.} \bibnamefont{Grant}} \bibnamefont{and}
  \bibinfo{author}{\bibfnamefont{H.~M.} \bibnamefont{Quiney}},
  \bibinfo{journal}{Adv. At. Mol. Phys.} \textbf{\bibinfo{volume}{23}},
  \bibinfo{pages}{37} (\bibinfo{year}{1988}).

\bibitem[{\citenamefont{Kotochigova and Tupitsin}(1987)}]{KotTup87}
\bibinfo{author}{\bibfnamefont{S.~A.} \bibnamefont{Kotochigova}}
  \bibnamefont{and} \bibinfo{author}{\bibfnamefont{I.~I.}
  \bibnamefont{Tupitsin}}, \bibinfo{journal}{J. Phys. B}
  \textbf{\bibinfo{volume}{20}}, \bibinfo{pages}{4759} (\bibinfo{year}{1987}).

\bibitem[{\citenamefont{J\"onsson and {\rm Froese Fischer}}(1994)}]{JonFro94}
\bibinfo{author}{\bibfnamefont{P.}~\bibnamefont{J\"onsson}} \bibnamefont{and}
  \bibinfo{author}{\bibfnamefont{C.}~\bibnamefont{{\rm Froese Fischer}}},
  \bibinfo{journal}{Phys. Rev. A} \textbf{\bibinfo{volume}{50}},
  \bibinfo{pages}{3080} (\bibinfo{year}{1994}).

\bibitem[{\citenamefont{Dzuba et~al.}(1996)\citenamefont{Dzuba, Flambaum, and
  Kozlov}}]{DzuFlaKoz96b}
\bibinfo{author}{\bibfnamefont{V.~A.} \bibnamefont{Dzuba}},
  \bibinfo{author}{\bibfnamefont{V.~V.} \bibnamefont{Flambaum}},
  \bibnamefont{and} \bibinfo{author}{\bibfnamefont{M.~G.}
  \bibnamefont{Kozlov}}, \bibinfo{journal}{Phys.\ Rev.\ A}
  \textbf{\bibinfo{volume}{54}}, \bibinfo{pages}{3948} (\bibinfo{year}{1996}).

\bibitem[{\citenamefont{Dzuba et~al.}(1998)\citenamefont{Dzuba, Kozlov, Porsev,
  and Flambaum}}]{DzuKozPor98}
\bibinfo{author}{\bibfnamefont{V.~A.} \bibnamefont{Dzuba}},
  \bibinfo{author}{\bibfnamefont{M.~G.} \bibnamefont{Kozlov}},
  \bibinfo{author}{\bibfnamefont{S.~G.} \bibnamefont{Porsev}},
  \bibnamefont{and} \bibinfo{author}{\bibfnamefont{V.~V.}
  \bibnamefont{Flambaum}}, \bibinfo{journal}{Zh. \ Eksp. \ Teor. \ Fiz.}
  \textbf{\bibinfo{volume}{114}}, \bibinfo{pages}{1636} (\bibinfo{year}{1998}),
  \bibinfo{note}{[Sov. \ Phys.--JETP {\bf 87} 885, (1998)]}.

\bibitem[{\citenamefont{Porsev et~al.}(1999{\natexlab{a}})\citenamefont{Porsev,
  Rakhlina, and Kozlov}}]{PorRakKoz99P}
\bibinfo{author}{\bibfnamefont{S.~G.} \bibnamefont{Porsev}},
  \bibinfo{author}{\bibfnamefont{{\rm Yu}.~G.} \bibnamefont{Rakhlina}},
  \bibnamefont{and} \bibinfo{author}{\bibfnamefont{M.~G.}
  \bibnamefont{Kozlov}}, \bibinfo{journal}{Phys. Rev. A}
  \textbf{\bibinfo{volume}{60}}, \bibinfo{pages}{2781}
  (\bibinfo{year}{1999}{\natexlab{a}}).

\bibitem[{\citenamefont{Porsev et~al.}(1999{\natexlab{b}})\citenamefont{Porsev,
  Rakhlina, and Kozlov}}]{PorRakKoz99J}
\bibinfo{author}{\bibfnamefont{S.~G.} \bibnamefont{Porsev}},
  \bibinfo{author}{\bibfnamefont{{\rm Yu}.~G.} \bibnamefont{Rakhlina}},
  \bibnamefont{and} \bibinfo{author}{\bibfnamefont{M.~G.}
  \bibnamefont{Kozlov}}, \bibinfo{journal}{J. Phys. B}
  \textbf{\bibinfo{volume}{32}}, \bibinfo{pages}{1113}
  (\bibinfo{year}{1999}{\natexlab{b}}).

\bibitem[{\citenamefont{Kolb et~al.}(1982)\citenamefont{Kolb, Johnson, and
  Shorer}}]{KolJohSho82}
\bibinfo{author}{\bibfnamefont{D.}~\bibnamefont{Kolb}},
  \bibinfo{author}{\bibfnamefont{W.~R.} \bibnamefont{Johnson}},
  \bibnamefont{and} \bibinfo{author}{\bibfnamefont{P.}~\bibnamefont{Shorer}},
  \bibinfo{journal}{Phys. Rev. A} \textbf{\bibinfo{volume}{26}},
  \bibinfo{pages}{19} (\bibinfo{year}{1982}).

\bibitem[{\citenamefont{Johnson et~al.}(1983)\citenamefont{Johnson, Kolb, and
  Huang}}]{JohKolHua83}
\bibinfo{author}{\bibfnamefont{W.~R.} \bibnamefont{Johnson}},
  \bibinfo{author}{\bibfnamefont{D.}~\bibnamefont{Kolb}}, \bibnamefont{and}
  \bibinfo{author}{\bibfnamefont{{\rm K.-N}.}~\bibnamefont{Huang}},
  \bibinfo{journal}{At. Data Nucl. Data Tables} \textbf{\bibinfo{volume}{28}},
  \bibinfo{pages}{333} (\bibinfo{year}{1983}).

\bibitem[{\citenamefont{Kozlov and Porsev}(1999)}]{KozPor99O}
\bibinfo{author}{\bibfnamefont{M.~G.} \bibnamefont{Kozlov}} \bibnamefont{and}
  \bibinfo{author}{\bibfnamefont{S.~G.} \bibnamefont{Porsev}},
  \bibinfo{journal}{Opt. Spectrosk.} \textbf{\bibinfo{volume}{87}},
  \bibinfo{pages}{384} (\bibinfo{year}{1999}), \bibinfo{note}{[Opt. \
  Spectrosc. {\bf 87}, 352 (1999)]}.

\bibitem[{\citenamefont{Brattsev et~al.}(1977)\citenamefont{Brattsev, Deineka,
  and Tupitsyn}}]{BraDeiTup77}
\bibinfo{author}{\bibfnamefont{V.~F.} \bibnamefont{Brattsev}},
  \bibinfo{author}{\bibfnamefont{G.~B.} \bibnamefont{Deineka}},
  \bibnamefont{and} \bibinfo{author}{\bibfnamefont{I.~I.}
  \bibnamefont{Tupitsyn}}, \bibinfo{journal}{Izv. Akad. Nauk SSSR, Ser. Fiz.}
  \textbf{\bibinfo{volume}{41}}, \bibinfo{pages}{2655} (\bibinfo{year}{1977}),
  \bibinfo{note}{[Bull. Acad. Sci. USSR, Phys. Ser. {\bf 41}, 173 (1977)]}.

\bibitem[{\citenamefont{Bogdanovich and \^Zukauskas}(1983)}]{BogZuk83}
\bibinfo{author}{\bibfnamefont{P.}~\bibnamefont{Bogdanovich}} \bibnamefont{and}
  \bibinfo{author}{\bibfnamefont{G.}~\bibnamefont{\^Zukauskas}},
  \bibinfo{journal}{Sov. Phys. Collect.} \textbf{\bibinfo{volume}{23}},
  \bibinfo{pages}{13} (\bibinfo{year}{1983}).

\bibitem[{\citenamefont{Moore}(1971)}]{Moo71}
\bibinfo{author}{\bibfnamefont{C.~E.} \bibnamefont{Moore}},
  \emph{\bibinfo{title}{Atomic energy levels}}, vol. \bibinfo{volume}{I-III}
  (\bibinfo{publisher}{NBS, National Standards Reference Data Series -- 35},
  \bibinfo{address}{U.S. GPO, Washington, D.C.}, \bibinfo{year}{1971}).

\bibitem[{\citenamefont{Oskay et~al.}(2006)}]{OskDid06}
\bibinfo{author}{\bibfnamefont{W. H.}~\bibnamefont{Oskay} \textit{et al}.},
  \bibinfo{journal}{Phys. Rev. Lett.} \textbf{\bibinfo{volume}{97}},
  \bibinfo{pages}{020801} (\bibinfo{year}{2006}).

\bibitem[{\citenamefont{Te et~al.}(2006)}]{TeJes03}
\bibinfo{author}{\bibfnamefont{Y.}~\bibnamefont{Te}},
  \bibinfo{journal}{Metrologia} \textbf{\bibinfo{volume}{40}},
  \bibinfo{pages}{24-30} (\bibinfo{year}{2003}).

\bibitem[{\citenamefont{Simon et~al.}(2003)}]{SimLau98}
\bibinfo{author}{\bibfnamefont{E.}~\bibnamefont{Simon}}
\bibinfo{author}{\bibfnamefont{P.}~\bibnamefont{Laurent}}
\bibnamefont{and}
  \bibinfo{author}{\bibfnamefont{A.}~\bibnamefont{Clairon}},
  \bibinfo{journal}{Phys. Rev. A} \textbf{\bibinfo{volume}{57}},
  \bibinfo{pages}{436-439} (\bibinfo{year}{1998}).

\bibitem[{\citenamefont{Cronin et~al.}(2007)}]{CroSch07}
\bibinfo{author}{\bibfnamefont{A.D.}~\bibnamefont{Cronin}}
\bibinfo{author}{\bibfnamefont{J.}~\bibnamefont{Schmiedmayer}}
\bibnamefont{and}
  \bibinfo{author}{\bibfnamefont{D. E.}~\bibnamefont{Pritchard}},
  \bibinfo{journal}{arXiv:0712.3703},
  \bibinfo{pages}{1-82} (\bibinfo{year}{2007}).

\bibitem[{\citenamefont{Ido and Katori}(2003)}]{IdoKat03}
\bibinfo{author}{\bibfnamefont{T.}~\bibnamefont{Ido}} \bibnamefont{and}
  \bibinfo{author}{\bibfnamefont{H.}~\bibnamefont{Katori}},
  \bibinfo{journal}{Phys. Rev. Lett.} \textbf{\bibinfo{volume}{91}},
  \bibinfo{pages}{053001} (\bibinfo{year}{2003}).

\bibitem[{\citenamefont{Derevianko et~al.}(1999)\citenamefont{Derevianko,
  Johnson, Safronova, and Babb}}]{DerJohSaf99}
\bibinfo{author}{\bibfnamefont{A.}~\bibnamefont{Derevianko}},
  \bibinfo{author}{\bibfnamefont{W.~R.} \bibnamefont{Johnson}},
  \bibinfo{author}{\bibfnamefont{M.~S.} \bibnamefont{Safronova}},
  \bibnamefont{and} \bibinfo{author}{\bibfnamefont{J.~F.} \bibnamefont{Babb}},
  \bibinfo{journal}{Phys.\ Rev.\ Lett.} \textbf{\bibinfo{volume}{82}},
  \bibinfo{pages}{3589} (\bibinfo{year}{1999}).

\bibitem[{\citenamefont{Porsev and Derevianko}(2003)}]{PorDer03}
\bibinfo{author}{\bibfnamefont{S.~G.} \bibnamefont{Porsev}} \bibnamefont{and}
  \bibinfo{author}{\bibfnamefont{A.}~\bibnamefont{Derevianko}},
  \bibinfo{journal}{J. Chem. Phys.} \textbf{\bibinfo{volume}{119}},
  \bibinfo{pages}{844} (\bibinfo{year}{2003}).

\bibitem[{\citenamefont{Kozlov et~al.}(1996)\citenamefont{Kozlov, Porsev, and
  Flambaum}}]{KozPorFla96}
\bibinfo{author}{\bibfnamefont{M.~G.} \bibnamefont{Kozlov}},
  \bibinfo{author}{\bibfnamefont{S.~G.} \bibnamefont{Porsev}},
  \bibnamefont{and} \bibinfo{author}{\bibfnamefont{V.~V.}
  \bibnamefont{Flambaum}}, \bibinfo{journal}{J. \ Phys. \ B}
  \textbf{\bibinfo{volume}{29}}, \bibinfo{pages}{689} (\bibinfo{year}{1996}).

\bibitem[{\citenamefont{Sternheimer}(1950)}]{Ste50}
\bibinfo{author}{\bibfnamefont{R.~M.} \bibnamefont{Sternheimer}},
  \bibinfo{journal}{Phys. Rev.} \textbf{\bibinfo{volume}{80}},
  \bibinfo{pages}{102} (\bibinfo{year}{1950}).

\bibitem[{\citenamefont{Dalgarno and Lewis}(1955)}]{DalLew55}
\bibinfo{author}{\bibfnamefont{A.}~\bibnamefont{Dalgarno}} \bibnamefont{and}
  \bibinfo{author}{\bibfnamefont{J.~T.} \bibnamefont{Lewis}},
  \bibinfo{journal}{Proc.\ Roy.\ Soc.} \textbf{\bibinfo{volume}{233}},
  \bibinfo{pages}{70} (\bibinfo{year}{1955}).


\bibitem[{\citenamefont{Sobelman}(1979)}]{Sob79}
\bibinfo{author}{\bibfnamefont{I.~I.} \bibnamefont{Sobelman}},
  \emph{\bibinfo{title}{Atomic Spectra And Radiative Transitions}}
  (\bibinfo{publisher}{Springer-Verlag}, \bibinfo{address}{Berlin, Heidelberg,
  New York}, \bibinfo{year}{1979}).

\bibitem[{\citenamefont{Werij et~al.}(1992)\citenamefont{Werij, Greene,
  Theodosiou, and Gallagher}}]{WerGreThe92}
\bibinfo{author}{\bibfnamefont{H.~{\rm G. C}.} \bibnamefont{Werij}},
  \bibinfo{author}{\bibfnamefont{C.~H.} \bibnamefont{Greene}},
  \bibinfo{author}{\bibfnamefont{C.~E.} \bibnamefont{Theodosiou}},
  \bibnamefont{and}
  \bibinfo{author}{\bibfnamefont{A.}~\bibnamefont{Gallagher}},
  \bibinfo{journal}{Phys. Rev. A} \textbf{\bibinfo{volume}{46}},
  \bibinfo{pages}{1248} (\bibinfo{year}{1992}).

\bibitem[{\citenamefont{Miller et~al.}(1992)\citenamefont{Miller, You, Cooper,
  and Gallagher}}]{MilYouCoo92}
\bibinfo{author}{\bibfnamefont{D.~A.} \bibnamefont{Miller}},
  \bibinfo{author}{\bibfnamefont{L.}~\bibnamefont{You}},
  \bibinfo{author}{\bibfnamefont{J.}~\bibnamefont{Cooper}}, \bibnamefont{and}
  \bibinfo{author}{\bibfnamefont{A.}~\bibnamefont{Gallagher}},
  \bibinfo{journal}{Phys. Rev. A} \textbf{\bibinfo{volume}{46}},
  \bibinfo{pages}{1303} (\bibinfo{year}{1992}).

\bibitem[{\citenamefont{Parkinson et~al.}(1976)\citenamefont{Parkinson, Reeves,
  and Tomkins}}]{ParReeTom76}
\bibinfo{author}{\bibfnamefont{W.~H.} \bibnamefont{Parkinson}},
  \bibinfo{author}{\bibfnamefont{E.~M.} \bibnamefont{Reeves}},
  \bibnamefont{and} \bibinfo{author}{\bibfnamefont{F.~S.}
  \bibnamefont{Tomkins}}, \bibinfo{journal}{J. Phys. B}
  \textbf{\bibinfo{volume}{9}}, \bibinfo{pages}{157} (\bibinfo{year}{1976}).

\bibitem[{\citenamefont{Nagel et~al.}(2005)\citenamefont{Nagel, Mickelson,
  Saenz, Martinez, Chen, Killian, Pellegrini, and C\^{o}t\'{e}}}]{NagMicSae05}
\bibinfo{author}{\bibfnamefont{S.~B.} \bibnamefont{Nagel}},
  \bibinfo{author}{\bibfnamefont{P.~G.} \bibnamefont{Mickelson}},
  \bibinfo{author}{\bibfnamefont{A.~D.} \bibnamefont{Saenz}},
  \bibinfo{author}{\bibfnamefont{Y.~N.} \bibnamefont{Martinez}},
  \bibinfo{author}{\bibfnamefont{Y.~C.} \bibnamefont{Chen}},
  \bibinfo{author}{\bibfnamefont{T.~C.} \bibnamefont{Killian}},
  \bibinfo{author}{\bibfnamefont{P.}~\bibnamefont{Pellegrini}},
  \bibnamefont{and}
  \bibinfo{author}{\bibfnamefont{R.}~\bibnamefont{C\^{o}t\'{e}}},
  \bibinfo{journal}{Phys. Rev. Lett.} \textbf{\bibinfo{volume}{94}},
  \bibinfo{pages}{083004} (\bibinfo{year}{2005}).

\bibitem{BorPenRed87}
E. N. Borisov, N. P. Penkin, and T. P. Redko, Opt. Spektrosk. {\bf
63} 475 (1987).

\bibitem{JonLevPer84}
G. J\"{o}nsson {\it et al.}, Z. Phys A {\bf 316}, 255 (1984).

\bibitem{Brinkmann69}
U. Brinkmann, Z. Phys. {\bf 228}, 449 (1969).

\bibitem{Havey77}
M. D. Havey, L. C. Balling, and J. J. Wright, J. Opt. Soc.
Am. {\bf 67}, 488 (1977).

\bibitem{Osherovich79}
A. L. Osherovich {\it et al.}, Opt. Spektrosk. {\bf 46} 243 (1979).

\bibitem{Andra75}
H. J. Andr\"{a} {\it et al.}, J. Opt. Soc. Am. {\bf 65}, 1410
(1975).

\bibitem{Gornik77}
W. Gornik, Z. Phys. A {\bf 283}, 231 (1977).

\bibitem{Fakuda}
K. Ueda, Y. Ashizawa, and K. Fukuda, J. Phys. Soc. Jpn. {\bf 51},
1936 (1982).

\end{thebibliography}

\end{document}